\documentclass[12pt]{article}
\usepackage{amsmath,amssymb, amsfonts, amsthm,amscd}
\usepackage[T2A]{fontenc}
\usepackage[cp1251]{inputenc}
\usepackage[russian,ukrainian,english]{babel}
\usepackage{geometry}
\usepackage{cite}

\newtheorem{Theorem}{Theorem}
\newtheorem{Corollary}{Corollary}
\newtheorem{Remark}{Remark}

\newtheorem{Proposition}{Proposition}

\newenvironment{TheoremProof}{\textbf{Proof. }}{\par\noindent\textbf{The Theorem is proved.}}
\newenvironment{PropositionProof}{\textbf{Proof. }}{\par\noindent\textbf{The Proposition is proved.}}

\title{{\Large \textbf{Some remarks on relations between the $\mu$-parameters of regular graphs}}}

\author{\normalsize N.N. Davtyan$^1$, R.R. Kamalian$^2$}

\date{\small{\small{$^1$Ijevan Branch of Yerevan State University,\\
e-mail: nndavtyan@gmail.com, \\
$^2$The Institute for Informatics and Automation Problems of NAS RA,\\
e-mail: rrkamalian@yahoo.com}}}

\begin{document}

\maketitle

\bigskip

\begin{abstract}
For an undirected, simple, finite, connected graph $G$, we denote by
$V(G)$ and $E(G)$ the sets of its vertices and edges, respectively.
A function $\varphi:E(G)\rightarrow \{1,...,t\}$ is called a proper
edge $t$-coloring of a graph $G$, if adjacent edges are colored
differently and each of $t$ colors is used. The least value of $t$
for which there exists a proper edge $t$-coloring of a graph $G$ is
denoted by $\chi'(G)$. For any graph $G$, and for any integer $t$
satisfying the inequality $\chi'(G)\leq t\leq |E(G)|$, we denote by
$\alpha(G,t)$ the set of all proper edge $t$-colorings of $G$. Let
us also define a set $\alpha(G)$ of all proper edge colorings of a
graph $G$:
$$
\alpha(G)\equiv\bigcup_{t=\chi'(G)}^{|E(G)|}\alpha(G,t).
$$

An arbitrary nonempty finite subset of consecutive integers is
called an interval. If $\varphi\in\alpha(G)$ and $x\in V(G)$, then
the set of colors of edges of $G$ which are incident with $x$ is
denoted by $S_G(x,\varphi)$ and is called a spectrum of the vertex
$x$ of the graph $G$ at the proper edge coloring $\varphi$. If $G$
is a graph and $\varphi\in\alpha(G)$, then define
$f_G(\varphi)\equiv|\{x\in V(G)/S_G(x,\varphi) \textrm{ is an
interval}\}|$.

For a graph $G$ and any integer $t$, satisfying the inequality
$\chi'(G)\leq t\leq |E(G)|$, we define:
$$
\mu_1(G,t)\equiv\min_{\varphi\in\alpha(G,t)}f_G(\varphi),\qquad
\mu_2(G,t)\equiv\max_{\varphi\in\alpha(G,t)}f_G(\varphi).
$$

For any graph $G$, we set:
$$
\mu_{11}(G)\equiv\min_{\chi'(G)\leq t\leq|E(G)|}\mu_1(G,t),\qquad
\mu_{12}(G)\equiv\max_{\chi'(G)\leq t\leq|E(G)|}\mu_1(G,t),
$$
$$
\mu_{21}(G)\equiv\min_{\chi'(G)\leq t\leq|E(G)|}\mu_2(G,t),\qquad
\mu_{22}(G)\equiv\max_{\chi'(G)\leq t\leq|E(G)|}\mu_2(G,t).
$$

For regular graphs, some relations between the $\mu$-parameters are
obtained.


Math. Classification: 05C15
\end{abstract}

We consider finite, undirected, connected graphs without loops and
multiple edges containing at least one edge. For any graph $G$, we
denote by $V(G)$ and $E(G)$ the sets of vertices and edges of $G$,
respectively. For any $x\in V(G)$, $d_G(x)$ denotes the degree of
the vertex $x$ in $G$. For a graph $G$, $\delta(G)$ and $\Delta(G)$
denote the minimum and maximum degrees of vertices in $G$,
respectively.

An arbitrary nonempty finite subset of consecutive integers is
called an interval. An interval with the minimum element $p$ and the
maximum element $q$ is denoted by $[p,q]$.

A function $\varphi:E(G)\rightarrow [1,t]$ is called a proper edge
$t$-coloring of a graph $G$, if each of $t$ colors is used, and
adjacent edges are colored differently.

The minimum value of $t$ for which there exists a proper edge
$t$-coloring of a graph $G$ is denoted by $\chi'(G)$ \cite{Vizing2}.

For any graph $G$, and for any $t\in[\chi'(G),|E(G)|]$, we denote by
$\alpha(G,t)$ the set of all proper edge $t$-colorings of $G$.

Let us also define a set $\alpha(G)$ of all proper edge colorings of
a graph $G$:
$$
\alpha(G)\equiv\bigcup_{t=\chi'(G)}^{|E(G)|}\alpha(G,t).
$$

If $\varphi\in\alpha(G)$ and $x\in V(G)$, then the set
$\{\varphi(e)/ e\in E(G), e \textrm{ is incident with } x$\} is
called a spectrum of the vertex $x$ of the graph $G$ at the proper
edge coloring $\varphi$ and is denoted by $S_G(x,\varphi)$.

If $G$ is a graph, $\varphi\in\alpha(G)$, then set
$V_{int}(G,\varphi)\equiv\{x\in V(G)/S_G(x,\varphi) \textrm{ is an
interval}\}$ and $f_G(\varphi)\equiv|V_{int}(G,\varphi)|$. A proper
edge coloring $\varphi\in\alpha(G)$ is called an interval edge
coloring \cite{Oranj3, Asratian4, Diss5} of the graph $G$ iff
$f_G(\varphi)=|V(G)|$. The set of all graphs having an interval edge
coloring is denoted by $\mathfrak{N}$. The terms and concepts which
are not defined can be found in \cite{West1}.

For a graph $G$, and for any $t\in[\chi'(G),|E(G)|]$, we set
\cite{Mebius6}:
$$
\mu_1(G,t)\equiv\min_{\varphi\in\alpha(G,t)}f_G(\varphi),\qquad
\mu_2(G,t)\equiv\max_{\varphi\in\alpha(G,t)}f_G(\varphi).
$$

For any graph $G$, we set \cite{Mebius6}:
$$
\mu_{11}(G)\equiv\min_{\chi'(G)\leq t\leq|E(G)|}\mu_1(G,t),\qquad
\mu_{12}(G)\equiv\max_{\chi'(G)\leq t\leq|E(G)|}\mu_1(G,t),
$$
$$
\mu_{21}(G)\equiv\min_{\chi'(G)\leq t\leq|E(G)|}\mu_2(G,t),\qquad
\mu_{22}(G)\equiv\max_{\chi'(G)\leq t\leq|E(G)|}\mu_2(G,t).
$$

Clearly, the $\mu$-parameters are correctly defined for an arbitrary
graph. Some remarks on their interpretations in games are given in
\cite{KornArxive, Petersen}.

The exact values of the parameters $\mu_{11}$, $\mu_{12}$,
$\mu_{21}$ and $\mu_{22}$ are found for simple paths, simple cycles
and simple cycles with a chord \cite{Simple7, Akunq}, "M\"{o}bius
ladders" \cite{Mebius6, Minchev}, complete graphs \cite{Arpine8},
complete bipartite graphs \cite{Arpine9, Arpine10}, prisms
\cite{Minchev, Arpine11}, $n$-dimensional cubes \cite{Arpine11,
Nikolaev12, KornArxive} and the Petersen graph \cite{Petersen}. The
exact values of $\mu_{11}$ and $\mu_{22}$ for trees are found in
\cite{Evg13}. The exact value of $\mu_{12}$ for an arbitrary tree is
found in \cite{Trees14} (see also \cite{Algorithm, Tree_Kontr}).

In this paper some relations between the $\mu$-parameters of regular
graphs are obtained.

In the rest part of this paper we admit an additional condition: an
arbitrary graph $G$ satisfies the inequality $\delta(G)\geq2$.

\begin{Theorem} \cite{Simple7, Akunq} \label{trm1}
For any integer $k\geq2$, the following equalities hold:
\begin{enumerate}
  \item $\mu_{12}(C_{2k})=\mu_{22}(C_{2k})=2k$,
  \item $\mu_{21}(C_{2k})=2k-1$,
  \item $\mu_{11}(C_{2k})=\left\{
\begin{array}{ll}
1, & \textrm{if $\;k=2$}\\
0, & \textrm{if $\;k\geq3$}
\end{array}
\right.$
\end{enumerate}
\end{Theorem}

\begin{Theorem} \cite{Simple7, Akunq} \label{trm2}
For any positive integer $k$, the following equalities hold:
\begin{enumerate}
  \item $\mu_{12}(C_{2k+1})=2$,
  \item $\mu_{21}(C_{2k+1})=\mu_{22}(C_{2k+1})=2k$,
  \item $\mu_{11}(C_{2k+1})=\left\{
\begin{array}{ll}
2, & \textrm{if $\;k=1$}\\
0, & \textrm{if $\;k\geq2$}
\end{array}
\right.$
\end{enumerate}
\end{Theorem}

\begin{Corollary} \cite{Simple7, Akunq} \label{cor1}
For any integer $k\geq2$, the inequalities
$\mu_{21}(C_{2k})<\mu_{12}(C_{2k})$ and
$\mu_{12}(C_{2k+1})<\mu_{21}(C_{2k+1})$ hold.
\end{Corollary}

\begin{Theorem} \cite{Simple7, Akunq} \label{trm3}
For any graph $G$, the inequalities
$\mu_{11}(G)\leq\mu_{12}(G)\leq\mu_{22}(G)$,
$\mu_{11}(G)\leq\mu_{21}(G)\leq\mu_{22}(G)$ hold.
\end{Theorem}

\begin{Remark} \cite{Simple7, Akunq} \label{rem1}
Corollary \ref{cor1} means that there are graphs $G$ for which
$\mu_{21}(G)<\mu_{12}(G)$ and there are also graphs $G$ for which
$\mu_{12}(G)<\mu_{21}(G)$.
\end{Remark}

\begin{Theorem} \cite{Simple7} \label{trm4}
If $G$ is a regular graph with $\chi'(G)=\Delta(G)$, then
$\mu_{12}(G)=|V(G)|$.
\end{Theorem}

\begin{Theorem} \cite{ArxivIneq} \label{trm5}
If $G$ is an $r$-regular graph, and $\varphi\in\alpha(G,|E(G)|)$,
then
$$
|V_{int}(G,\varphi)|\leq\bigg\lfloor\frac{r\cdot|V(G)|-2}{2\cdot(r-1)}\bigg\rfloor.
$$
\end{Theorem}

\begin{Corollary}\label{cor2}
If $G$ is an $r$-regular graph, then
$$
\mu_2(G,|E(G)|)\leq\bigg\lfloor\frac{r\cdot|V(G)|-2}{2\cdot(r-1)}\bigg\rfloor.
$$
\end{Corollary}

\begin{Corollary}\label{cor3}
If $G$ is an $r$-regular graph, then
$$
\mu_{21}(G)\leq\bigg\lfloor\frac{r\cdot|V(G)|-2}{2\cdot(r-1)}\bigg\rfloor
$$
\end{Corollary}

\begin{Proposition}\label{prop1}
For arbitrary integers $r\geq2$ and $n\geq1$, the inequality
$$
\bigg\lfloor\frac{r\cdot n-2}{2\cdot(r-1)}\bigg\rfloor\leq n-1
$$
holds.
\end{Proposition}

\begin{PropositionProof}
$$
\bigg\lfloor\frac{rn-2}{2\cdot(r-1)}\bigg\rfloor=\bigg\lfloor\frac{n}{2}+
\frac{n-2}{2\cdot(r-1)}\bigg\rfloor\leq\bigg\lfloor\frac{n}{2}+
\frac{n-2}{2}\bigg\rfloor=n-1.
$$
\end{PropositionProof}

\begin{Corollary}\label{cor4}
If $G$ is a regular graph, then $\mu_{21}(G)\leq|V(G)|-1$.
\end{Corollary}

From corollary \ref{cor4} and theorem \ref{trm4} we obtain

\begin{Corollary}\label{cor5}
For an arbitrary regular graph $G$ with $\chi'(G)=\Delta(G)$, the
inequality $\mu_{21}(G)<\mu_{12}(G)$ holds.
\end{Corollary}

\begin{Theorem} \label{trm6}
For an arbitrary regular graph $G$, the following four statements
are equivalent:
\begin{enumerate}
  \item $\chi'(G)=\Delta(G)$,
  \item $G\in\mathfrak{N}$,
  \item $\mu_{22}(G)=|V(G)|$,
  \item $\mu_{12}(G)=|V(G)|$.
\end{enumerate}
\end{Theorem}

\begin{TheoremProof}
The equivalence between 1) and 2) was proved in \cite{Oranj3,
Asratian4, Diss5}. The equivalence between 2) and 3) is evident.

Let us show the equivalence between 1) and 4).

If $\chi'(G)=\Delta(G)$, then, by theorem \ref{trm4}, we have the
equality $\mu_{12}(G)=|V(G)|$. It means that $1)\Rightarrow 4)$.

Now suppose that $\mu_{12}(G)=|V(G)|$. By theorem \ref{trm3}, we
have also the equality $\mu_{22}(G)=|V(G)|$. Consequently, using the
equivalence between 2) and 3), we have also the relation
$G\in\mathfrak{N}$. Finally, using the equivalence between 1) and
2), we have also the equality $\chi'(G)=\Delta(G)$. Thus,
$4)\Rightarrow 1)$.
\end{TheoremProof}


\bigskip

\begin{thebibliography}{1}

\bibitem{Vizing2}  {V.G.~Vizing},
                {\it The chromatic index of a multigraph}, Kibernetika 3 (1965),
                pp. 29--39.

\bibitem{Oranj3} {A.S.~Asratian, R.R.~Kamalian},
                {\it Interval colorings of edges of a multigraph}, Appl. Math. 5 (1987),
               Yerevan State University, pp. 25--34 (in Russian).

\bibitem{Asratian4} {A.S.~Asratian, R.R.~Kamalian},
                {\it Investigation of interval edge-colorings of graphs},
                Journal of Combinatorial Theory. Series B 62 (1994),
                no.1, pp. 34--43.

\bibitem{Diss5} {R.R.~Kamalian}, {\it Interval Edge Colorings of Graphs},
                Doctoral dissertation, the Institute of Mathematics of the Siberian Branch of
                the Academy of Sciences of USSR, Novosibirsk, 1990 (in Russian).

\bibitem{West1}  {D.B.~West}, {\it Introduction to Graph Theory}, Prentice-Hall, New Jersey, 1996.

\bibitem{Mebius6}  {N.N.~Davtyan, R.R.~Kamalian},
                {\it On boundaries of extremums of the number of vertices with an interval spectrum among the set of
                proper edge colorings of "M\"{o}bius ladders" with $t$ colors under variation of $t$}, Proc. of the
                $3^{th}$ Ann. Sci. Conf. (December 5--10, 2008) of the RAU, Yerevan, 2009, pp. 81--84 (in Russian).

\bibitem{KornArxive}   {A.M.~Khachatryan, R.R.~Kamalian},
                {\it On the extremal values of the number of vertices with an interval spectrum on the set of proper edge colorings of the graph
                of the $n$-dimensional cube}, http://arxiv.org/abs/1307.1389

\bibitem{Petersen}  {N.N.~Davtyan},
                {\it On the $\mu$-parameters of the Petersen graph}, http://arxiv.org/abs/1307.2348

\bibitem{Simple7} {N.N.~Davtyan, R.R.~Kamalian}, {\it On properties of the number of vertices with an interval spectrum in
                proper edge colorings of some graphs}, the Herald of the RAU,  №2, Yerevan, 2009, pp. 33--42.

\bibitem{Akunq} {N.N.~Davtyan, R.R.~Kamalian}, {\it Some properties of the number of vertices with an interval spectrum in
                proper edge colorings of graphs}, The Collection "Akunq" of Scientific Papers of Ijevan Branch of Yerevan State University,  Yerevan, 2012, pp. 18--27.

\bibitem{Minchev}  {N.N.~Davtyan, A.M.~Khachatryan, R.R.~Kamalian},
                {\it On Boundaries of Extrema of the Number of Vertices with an Interval Spectrum on the Sets of Proper Edge $t$-colorings of Some Cubic Graphs under Variation of $t$},
                International Mathematical Forum, Vol. 8, 2013, no. 24, pp. 1195--1198, http://dx.doi.org/10.12988/imf.2013.3491.

\bibitem{Arpine8} {A.M.~Khachatryan}, {\it On boundaries of extremums of the number of vertices with an interval spectrum among the set of
                proper edge colorings of complete graphs with $t$ colors under variation of $t$}, Proc. of the
                $5^{th}$ Ann. Sci. Conf. (December 6--10, 2010) of the RAU, Yerevan, 2011, pp. 268--272 (in Russian).

\bibitem{Arpine9} {A.M.~Khachatryan}, {\it On the parameters $\mu_{11}$, $\mu_{12}$ and $\mu_{22}$ of complete bipartite graphs},
                the Herald of the RAU,  №1, Yerevan, 2011, pp. 76--83 (in Russian).

\bibitem{Arpine10} {R.R.~Kamalian, A.M.~Khachatryan}, {\it On the sharp value of the parameter $\mu_{21}$ of complete bipartite graphs},
               the Herald of the RAU,  №2, Yerevan, 2011, pp. 19--25 (in Russian).

\bibitem{Arpine11} {R.R.~Kamalian, A.M.~Khachatryan}, {\it On properties of a number of vertices with an interval spectrum among the set
                of proper edge colorings of some regular graphs}, Proc. of the $6^{th}$ Ann. Sci. Conf. (December 5--9, 2011) of the RAU,
                Yerevan, 2012, pp. 62--65 (in Russian).

\bibitem{Nikolaev12}   {A.M.~Khachatryan, R.R.~Kamalian},
                {\it On the $\mu$-parameters of the graph of the $n$-dimensional cube}, Book of abstracts of the International Mathematical
                Conference on occasion to the 70th year anniversary of Professor Vladimir Kirichenko, June 13-19 (2012), Mykolaiv, Ukraine, pp. 38--39.

\bibitem{Evg13}   {N.N.~Davtyan},
                {\it On the least and the greatest possible numbers of vertices with an interval spectrum on the set of
                proper edge colorings of a tree}, Math. Problems of Computer Science, Vol. 32, Yerevan, 2009, pp. 107--111 (in Russian).

\bibitem{Trees14} {N.N.~Davtyan, R.R.~Kamalian}, {\it On the parameter $\mu_{12}$ of a tree}, Proc. of the $4^{th}$
                Ann. Sci. Conf. (November 30 -- December 4, 2009) of the RAU, Yerevan, 2010, pp. 149--151 (in Russian).

\bibitem{Algorithm} {N.N.~Davtyan, R.R.~Kamalian}, {\it On an algorithm of evaluation of the exact value of the parameter
               $\mu_{12}$ of an arbitrary tree}, the Herald of the RAU, №1, Yerevan, 2011, pp. 57--63 (in Russian).

\bibitem{Tree_Kontr} {N.N.~Davtyan}, {\it On a property of the parameter $\mu_{12}$ of trees of special kind}, the Herald of the RAU, №2, Yerevan, 2010, pp. 77--82 (in Russian).

\bibitem{ArxivIneq}  {N.N.~Davtyan, R.R. Kamalian},
                {\it An inequality for the number of vertices with an interval spectrum in edge labelings of regular graphs},
                http://arxiv.org/abs/1307.1392.
\end{thebibliography}
\end{document}